\documentstyle[psfig]{l-aa}

\begin{document}

   \thesaurus{07          
              (08.01.1;   
               08.05.3;   
               08.16.4;   
               08.09.2: V854\,Cen;   
               08.09.2: Sakurai's object;   
               08.22.3)}   

   \title{Abundance similarities between the 
R\,CrB star V854\,Cen and 
the born-again 
Sakurai's object}

 \author{Martin Asplund\inst{1,2}, Bengt Gustafsson\inst{1}, 
N. Kameswara Rao\inst{3} and David L. Lambert\inst{4, }
\thanks{Visiting Astronomer at Cerro Tololo Inter-American
              Observatory (CTIO), which is operated by the
              Association of Universities for Research in Astronomy Inc.,
              under contract with the National Science Foundation}}

   \offprints{M. Asplund (martin@nordita.dk)}

   \institute{              
              Astronomiska observatoriet,
              Box 515,
              S--751 20 ~Uppsala,
              Sweden\\
              \and
              present address: NORDITA, 
              Blegdamsvej 17, 
              DK-2100 ~Copenhagen {\O}, 
              Denmark\\              
              \and
              Indian Institute of Astrophysics,
              Bangalore 560 034,
              India \\
              \and
              Department of Astronomy,
              University of Texas,
              Austin, TX 78712,
              USA \\
              }

   \date{
Accepted for publication in {\it Astronomy \& Astrophysics}}

   \maketitle

   \markboth
{Asplund et al.: Abundance similarities between 
V854\,Cen and Sakurai's object}
{Asplund et al.: Abundance similarities between 
V854\,Cen and Sakurai's object}

\begin{abstract}                                                                           
The elemental abundances of the mildly hydrogen-deficient
R Coronae Borealis (R\,CrB) star V854\,Cen 
have been estimated.
The R\,CrB stars have been divided into
majority and  minority classes judging by their
abundance patterns. Class assignment has previously been unambiguous
but V854\,Cen has traits of both the
minority and majority class. 
Neither V854\,Cen nor the three obvious minority members  show
any clear abundance 
signatures of having been affected by e.g. dust-gas separation
as often observed in post-AGB stars.
By chemical composition, V854\,Cen closely resembles
Sakurai's object, which has probably recently experienced
a final He-shell flash. 
Therefore V854\,Cen and Sakurai's object may share the same 
evolutionary background, which would add 
support for the final-flash scenario
as a viable origin of the R\,CrB stars.
Most of the few differences in abundance ratios 
between the stars 
could if so be attributed to milder H-ingestion in
connection with the final He-shell
flash of V854\,Cen.
The identification of either the majority or the minority
group, if any, 
as final flash objects, remain uncertain, however, due to the 
unclear membership status of V854\,Cen.


\end{abstract}

\section{Introduction}

The extreme hydrogen-deficiency
of the R Coronae Borealis (R\,CrB) stars together with  
overabundances of elements produced at the time of 
helium burning suggest that these stars
are in late evolutionary stages.
A removal of essentially all of the 
hydrogen-rich envelope prior to leaving the asymptotic
giant branch (AGB) is very unlikely in the context of
single star evolution for low- and intermediate-mass stars.
Furthermore, due to the short time-scale 
following evolution off the AGB and the lack of significant
pre-planetary nebula material, the stars cannot readily
be identified with a phase immediately after departing from 
the AGB. Instead, the R\,CrB stars are likely to be re-born stars,
i.e., stars that evolved off the AGB to the white dwarf 
cooling regime but now show brief
re-appearances as luminous giants.

Two proposals for re-born stars have garnered most attention:
the ``final flash'' (Renzini 1979) and the ``double 
degenerate'' (Webbink 1984) models. 
In the former a final He-shell flash in a post-AGB star
descending the white dwarf cooling track briefly expands the stellar
envelope to giant dimensions once again.
This He-shell flash quickly
depletes the envelope of hydrogen to create an R\,CrB-like supergiant.
Examples of such stars may be the recently discovered Sakurai's object
(Asplund et al. 1997b) and FG\,Sge (Gonzalez et al. 1998).
The second model involves a merger of a He white dwarf and
a C-O white dwarf.
Close white dwarf binaries such as WD\,2331+290 and WD\,0957-666 
may merge within a Hubble time 
and produce hydrogen-deficient giants (Iben et al. 1997).
The pros and cons of each scenario are
discussed in Iben et al. (1996) and Sch\"onberner (1996).
Possible progenitors and descendants to the R\,CrB stars
are the hydrogen-deficient carbon (HdC) stars (Warner 1967)
and the extreme helium (EHe) stars (e.g. Jeffery 1996),
respectively, which may subsequently evolve into
the helium sub-dwarf O (HesdO$^+$) stars.

The two suggested scenarios will probably produce different
atmospheric chemical compositions.
Estimates of the elemental abundances 
are therefore essential for placing the stars in the correct
context of stellar evolution.
In two companion articles the compositions of 17 R\,CrB stars
and two EHe stars (Lambert et al. 1998) and Sakurai's object 
(Asplund et al. 1997b) have been determined.
V854\,Cen is discussed separately being a `peculiar' (!) R\,CrB star
with two special characteristics: a relatively high
hydrogen abundance and a propensity to undergo frequent declines.
In the present paper the composition of V854\,Cen is analysed. 
Similarities and differences with the compositions of R\,CrB 
stars and Sakurai's object are investigated and discussed 
in light of the proposed evolutionary scenarios.
  
\section{Observations}

V854\,Cen was observed in July 1989 with the 
Cassegrain echelle spectrograph on the CTIO 4m reflector,
together with many other R\,CrB stars which form the
basis for the analysis described in Lambert et al. (1998).
At the time, the star was at or close to maximum light.  
Both red (5500-6840\,\AA) and blue (4200-4900\,\AA) spectra
were obtained at a resolution of about 0.3\,\AA.
The signal-to-noise of the different spectra
is 100 or higher over most of the observed bandpass.

For the C$_2$ Swan 0-1 band, the He\,{\sc i} triplet 
and the hydrogen lines, 
synthetic spectroscopy were used to fit the spectral features.
Otherwise, equivalent widths of seemingly unblended 
lines were measured. A list of all lines used for the analysis
can be found in Table \ref{t:lines}. 

\section{Abundance analysis}

The abundance analysis of V854\,Cen is based on 
line-blanketed, hydrogen-deficient model atmospheres similar to those
described in Asplund et al. (1997a).
The input abundances for the model atmosphere calculations were 
determined in an iterative fashion, such that all abundances of the 
final adopted model were
consistent with the results from the abundance analysis, with one important
exception, namely carbon, which will be discussed further below.

\begin{figure}[t]
\centerline{
\psfig{figure=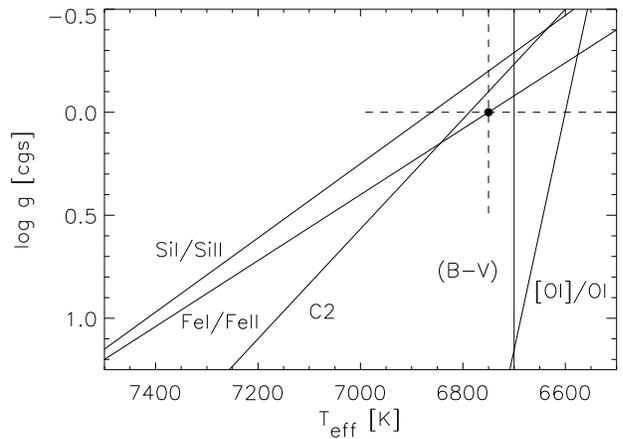,width=9cm}}
\caption{The loci provided by the different $T_{\rm eff}$-log\,$g$ 
indicators for C/He=10\%. The adopted stellar parameters 
$T_{\rm eff} = 6750$\,K and log\,$g = 0.0$\,[cgs] are denoted by 
$\bullet$, with the estimated uncertainties shown by dashed lines. 
Consistent parameters which are not shown in the figure 
were also obtained from the excitation balance
of Fe\,{\sc i} and {\sc ii} lines and the hydrogen lines}
         \label{f:param}
\end{figure}

The stellar parameters $T_{\rm eff} = 6750\pm 250$\,K, 
log\,$g =0.0 \pm 0.5$ and 
$\xi_{\rm turb}=6.0\pm 1.0 \,$km\,s$^{-1}$ 
were determined utilizing a range of spectral features. 
The ionization balances of Fe\,{\sc i}/Fe\,{\sc ii}
and Si\,{\sc i}/Si\,{\sc ii} provide loci in the $T_{\rm eff}$-log\,$g$ 
diagram, which run essentially parallel with the solution from 
the C$_2$ Swan band strengths. The excitation balance from
high-excitation, permitted O\,{\sc i} lines and low-excitation forbidden
[O\,{\sc i}] lines provide an additional temperature indicator with a minor
gravity dependence.
Also the excitation balance of Fe\,{\sc i} and Fe\,{\sc ii} lines and
$B-V$ photometry (adopting $(B-V)_0 = 0.50$, N.K. Rao 
unpublished research) provide estimates of $T_{\rm eff}$.
V854\,Cen has only a small light amplitude due to pulsations
and the non-simultaneous photometry compared with the spectroscopy
is therefore unlikely to introduce any significant errors. 
The loci in the $T_{\rm eff}$-log\,$g$ diagram
of the various indicators are shown in Fig. \ref{f:param}.
The microturbulence parameter $\xi_{\rm turb}$
 was derived from Fe\,{\sc ii} 
and Ti\,{\sc ii} lines of various strengths.

With the estimated $T_{\rm eff}$ and log\,$g$, the H abundance
was determined using the wings of H$\beta$. 
H$\gamma$ was also used but is
more affected by blending and was therefore given a lower weight, though
the derived abundance was consistent with the
value obtained from H$\beta$.
Synthetic spectra of H$\alpha$ was not attempted since
the line is often distorted by emission components even
at maximum light for R\,CrB stars (Rao \& Lambert 1997).
For the computed H line profiles, data were taken from Seaton (1990
and private communication).
The weak Balmer lines in V854\,Cen, compared 
to normal supergiants of similar 
temperature, require a significant H-deficiency, since the 
strengths of the H line wings remain unchanged or even 
{\it increase} with a decreasing 
H abundance until another element but H 
takes over as the dominant opacity
source (B\"ohm-Vitense 1979).
The Balmer line profiles give 
additional information on $T_{\rm eff}$ and log\,$g$, 
which for V854\,Cen were consistent with those of 
the other indicators. 

The continuous opacity at visual wavelengths in atmospheres of R\,CrB
stars is predicted to be provided largely by photoionization of
C\,{\sc i} from highly excited levels (Searle 1961; Asplund et al. 1997a). 
Then, the equivalent widths
of C\,{\sc i} lines from similarly excited levels must be almost
independent of the atmospheric parameters $T_{\rm eff}$ and log\,$g$. 
This expectation is confirmed by the fact that the equivalent 
widths of weak C\,{\sc i} lines are almost the same from one 
R\,CrB star to another.
However, the analyses of the R\,CrB stars has 
revealed a significant discrepancy between the 
observed and predicted line strengths of 
the C\,{\sc i} lines (Gustafsson \& Asplund 1996; Lambert et al. 1998):
all observed lines are weaker than predicted to the extent 
that the derived C abundance is on average 0.6\,dex 
less than the input abundance 
with a 
small scatter between the stars. 
The reason for this so called
``carbon problem'' is still unknown, but it may be related to
inappropriate assumptions for the model atmospheres (one-dimensional,
static, flux constant models in LTE) on which the analyses
are based (Lambert et al. 1998). 
Fortunately, various tests suggest 
that abundance ratios such as [X/Fe]
\footnote{The abundance ratios are defined by the customary
[X/Fe]=log\,(X/Fe)$_* - $log\,(X/Fe)$_\odot$}  
will in general be little affected by the 
carbon problem.
As in the previous studies, we have chosen to 
adopt a C abundance for the model atmospheres, 
which results in a ``carbon problem'' of 0.8\,dex.

Despite the fact that C dominates the continuous opacity, 
He is the most abundant element.
In principle, the He abundance may be determined from 
the presence of the He\,{\sc i} 5876\,\AA\
triplet.
The predicted He\,{\sc i} D$_3$ triplet 
(with the blending C\,{\sc i}
lines taken into account using synthetic spectroscopy,
Lambert et al. 1998) 
is too strong 
with an assumed C/He ratio of 1\% by number, which is the value 
determined for most of the EHe stars (Jeffery 1996) 
and deemed appropriate for the 
R\,CrB stars by Lambert et al. (1998).
A ratio of 10\% seems more suitable,
which is the value found for 
Sakurai's object (Asplund et al. 1997b).
The C/He ratio in V854\,Cen is uncertain, however, 
due to the significant
contribution of the blending C\,{\sc i} lines,
and possible departures from LTE for this high-excitation
transition.
The derived stellar parameters are independent to first order of 
the C/He ratio, 
since He is acting only as an inert element, only
contributing to the gas pressure except for at large depths.
The effects of the uncertainty in C/He will thus be minor;
typically, the [X/Fe] ratios differ by $<0.05$\,dex when using 1\% 
instead of 10\% for the C/He ratio, though
smaller absolute abundances 
will be derived,
as is clear from Table \ref{t:abund}.

A list of the lines used for the abundance analysis with 
relevant data is shown in Table \ref{t:lines}. 
The adopted {\it gf}-values are the 
same as in Lambert et al. (1998),
i.e. mainly from a compilation kindly provided by R.E. Luck.
For the Fe\,{\sc i} and {\sc ii} lines, data were 
taken from Lambert et al. (1996).
For the C\,{\sc i} lines, Opacity Project {\it gf}-values 
(Seaton et al. 1994; Luo \& Pradhan 1989) 
were used assuming LS-coupling, 
as well as data from Hibbert et al. (1993)
for intermediate coupling.
Most of the lines used for abundance determination 
are weak and hence little affected by microturbulence
and hyperfine structure, except for Mg, Sc, Ti, Cr, Sr and Ba.
For two of the Ba\,{\sc ii} lines we have investigated the
effects of hyperfine structure, with relevant data 
taken from Magain (1995)
for $\lambda 4554.04$\,\AA $ $ and Villemoes (1993) 
for $\lambda 6496.90$\,\AA . Isotope shifts were not included,
since they are expected to play a minor role (Cowley \& Frey 1989). 
The hyperfine structure affects
the derived abundances from the lines by only $\leq 0.02$\,dex,
due to the fact that $\xi_{\rm turb}$ is large relative 
to the hyperfine splitting.
Thus, the neglect of hyperfine structure will probably not 
compromise our conclusions regarding the chemical composition.

\section{Chemical composition and comparison with the other R\,CrB stars}

\subsection{Elemental abundances}

\begin{table*}[t]
\caption{ Chemical compositions of V854\,Cen, the 
R\,CrB stars, Sakurai's object and the Sun.
The customary normalization 
log\,($\Sigma \mu_i \epsilon_i$) = 12.15 has been
adopted for the abundances
\label{t:abund}
}
\begin{tabular}{lccccccccccc} 
\\
 \hline \\

 Element & Sun$^{\rm a}$ & \multicolumn {2} {c} {V854\,Cen} & &
\multicolumn {2} {c} {Sakurai's object$^{\rm b}$} & 
R\,CrB & \multicolumn {4} {c} {R\,CrB minority}  \\ 
 \cline{3-4} \cline{6-7}  \cline{9-12} \\
 && C/He=1\% & C/He=10\% && May & October & majority$^{\rm c}$ 
& V\,CrA$^{\rm c}$ & V3795\,Sgr$^{\rm c}$  & VZ\,Sgr$^{\rm c}$ & DY\,Cen$^{\rm d}$  \\            
\hline \\ 
H  & 12.00 & $8.9\pm 0.1$ & $9.9\pm 0.1$ && 9.7 & 9.0 &   &   8.0 &  $<4.1$ &  6.2 & 10.8  \\
He & 10.99 & 11.5 & 11.4 &&
11.4$^{\rm e}$ & 11.4$^{\rm e}$ & 11.5$^{\rm e}$ & 11.5$^{\rm e}$ & 11.5$^{\rm e}$
 & 11.5$^{\rm e}$ & 11.5   \\
Li &  3.31$^{\rm a}$ &  $<1.0$ & $<2.0$ && 3.6 & 4.2 &       &  & &  & \\ 
C  &  8.55 &  $8.7^{\rm f}\pm 0.3$ & $9.6^{\rm f}\pm 0.3$ && 9.7$^{\rm f}$ 
& 9.8$^{\rm f}$ & 8.9$^{\rm f}$ & 8.6$^{\rm f}$ & 8.8$^{\rm f}$ &  8.8$^{\rm f}$ & 9.5 \\
N  &  7.97 &  $6.8\pm 0.1$ & $7.8\pm 0.1$ && 8.9 &  8.9 &  8.6 & 8.6 & 8.0 & 7.6 & 8.0 \\  
O  &  8.87 &  $7.9\pm 0.1$ & $8.9\pm 0.1$ && 9.5 &  9.4 &  8.2 & 8.7 & 7.5 & 8.7 & 8.8 \\
Na &  6.33 &  $5.4\pm 0.1$ & $6.4\pm 0.1$ && 6.7 &  6.8 &  6.1 & 5.9 & 5.9 & 5.8 & \\
Mg &  7.58 &  $5.2$~~~~~~~~ & $6.2$~~~~~~~~ && 6.6 &  6.5 & 6.4 & 6.6 & 6.1 &     & 7.3 \\
Al &  6.47 &  $4.7\pm 0.1$ & $5.7\pm 0.1$ && 6.6 &  6.3 &  6.0 & 5.3 & 5.6 & 5.4 & 5.9 \\  
Si &  7.55 &  $6.1\pm 0.2$ & $7.0\pm 0.2$ && 7.1 &  7.5 &  7.1 & 7.6 & 7.5 & 7.3 & 8.1 \\ 
S  &  7.23 &  $5.5\pm 0.1$ & $6.4\pm 0.1$ && 6.6 &  6.9 &  6.9 & 7.6 & 7.4 & 6.7 & 7.1 \\  
Ca &  6.36 &  $4.2\pm 0.2$ & $5.1\pm 0.2$ && 5.6 &  5.5 &  5.4 & 5.1 & 5.3 & 5.0 & \\
Sc &  3.17 &  $1.9\pm 0.1$ & $2.9\pm 0.1$ && 3.1 &  3.9 &  2.7 & 2.8 & 2.8 \\  
Ti &  5.02 &  $3.1\pm 0.2$ & $4.1\pm 0.2$ && 4.1 &  4.6 &  4.0 & 3.3 & 3.5 \\  
Cr &  4.00 &  $3.2\pm 0.2$ & $4.2\pm 0.2$ && 4.5 &  5.1 &      &     & 4.2 \\  
Fe &  7.50 &  $5.0\pm 0.1$ & $6.0\pm 0.1$ && 6.3 &  6.6 &  6.5 & 5.4 & 5.6 & 5.8 & 5.0 \\  
Ni &  6.25 &  $4.9\pm 0.1$ & $5.9\pm 0.1$ && 6.1 &  6.2 &  5.9 & 5.2 & 5.8 & 5.2 & \\  
Zn &  4.60 &  $3.4\pm 0.3$ & $4.4\pm 0.3$ && 4.7 &  5.4 &  4.4 & 4.0 & 4.1 & 3.8 & \\ 
Sr &  2.97 &  $1.2$~~~~~~~~ & $2.2$~~~~~~~~ && 4.9 &  5.4: &      &            \\
Y  &  2.24 &  $1.2\pm 0.2$ & $2.2\pm 0.2$ && 3.3 &  4.2 &  2.1 & 0.6 & 1.5 & 2.8 & \\     
Zr &  2.60 &  $1.1\pm 0.2$ & $2.1\pm 0.2$ && 3.0 &  3.5 &  2.0 &     &     & 2.5 &  \\
Ba &  2.13 &  $0.3\pm 0.1$ & $1.3\pm 0.1$ && 1.5 &  1.9 &  1.6 & 0.7 & 0.9 & 1.4 &\\   
La &  1.17 &  $-0.6\pm 0.1$ & $0.4\pm 0.1$ && $<1.6$ &  1.5 &      &  \\
Ce &  1.58 &  $-0.5\pm 0.2$ & $0.5\pm 0.2$ &&  &   &      &  \\
\hline  \\
\end{tabular}

\begin{list}{}{}
\item[$^{\rm a}$] From Grevesse et al. (1996). For Li the meteoritic 
value is adopted.
\item[$^{\rm b}$] From Asplund et al. (1997b)
\item[$^{\rm c}$] From Lambert et al. (1997).
The majority is an average of 14 stars, with little scatter 
except for O, Y and Ba. For Mg, Sc, Ti and Zr the mean is based only
on a few majority stars
\item[$^{\rm d}$] From Jeffery \& Heber (1993). 
\item[$^{\rm e}$] Input C/He ratio for model atmospheres: 10\% estimated for Sakurai's object
and C/He=1\% assumed for the R\,CrB stars, except for DY\,Cen where
it was determined spectroscopically to be 1\%.
\item[$^{\rm f}$] Spectroscopically determined C\,{\sc i} abundance, which differs 
from assumed input abundance by typically 0.6\,dex.
\end{list}

\end{table*}

The derived abundances from the LTE analysis are listed 
in Table \ref{t:abund}.
The given errors are the formal standard deviations 
from the different lines; for elements with only a single line, 
no error is given.
Errors introduced by the uncertainties in the adopted stellar parameters
are given in Table \ref{t:error}, from which we judge the accuracy
of most determined abundances to be better than 
$0.3$\,dex, except for Ca, Ni, Zn, Sr and Ba where
the errors may be slightly larger.
Abundance ratios such as [X/Fe] will in general 
be much less vulnerable to errors in the parameters.

\begin{figure}[t]
\centerline{
\psfig{figure=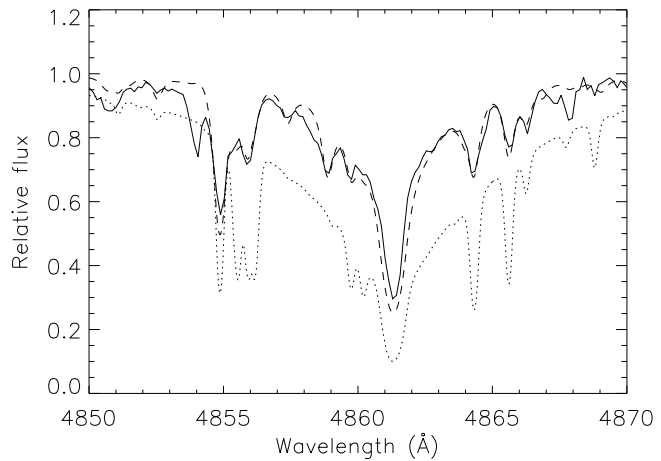,height=6.8cm}}
\caption{H$\beta$ in V854\,Cen (solid) compared with synthetic spectra
with normal hydrogen abundance (dots) and H-deficient by 2.1\,dex (dashed). 
The stellar parameters used 
for the predicted profiles are $T_{\rm eff} = 6750$\,K,
log\,$g = 0.00$ [cgs] and $\xi_{\rm turb} = 6.0$\,km\,s$^{-1}$.
For the H-deficient model C/He=10\% has been used}
         \label{f:hbeta}
\end{figure}

According to Table \ref{t:abund} V854\,Cen is metal-poor. 
Assuming C/He=10\%, the Fe mass fraction
is 0.6\,dex below solar if the spectroscopic C abundance is adopted
(1.4\,dex with the input C abundance).
In fact, the Fe/C ratio is the lowest of all analysed 
R\,CrB stars with the exception of DY\,Cen.
Had the usual C/He=1\% ratio (Lambert et al. 1998) 
been assumed, the derived metallicity might have been problematic
considering the galactic location of the star:
Z\,$=800$\,pc (assuming $M_{\rm bol} = -5$, 
Lambert et al. 1998).
The latter suggests that it may belong to the thick disk 
population, and thus not metal-poor by a large factor.
If C/He=1\% is still to be preferred, V854\,Cen may have
acquired the metal-poorness (then 1.6\,dex below solar) 
through chemical processes.
One chemical process -- the separation of dust and gas --
will be investigated in detail below. 

\begin{table}[t]
\caption{Abundance errors due to uncertainties in the stellar parameters of
V854\,Cen, defined by $\Delta$(log\,$\epsilon_i$) = log\,$\epsilon_i$(perturbed) -
log\,$\epsilon_i$(adopted). 
The adopted parameters are 
$T_{\rm eff} = 6750$\,K, log\,$g = 0.00$\,[cgs] and 
$\xi_{\rm turb} = 6.0$\,km\,s$^{-1}$
\label{t:error}
}
\begin{tabular}{lccc} 
\\
 \hline \\
  Element & $\Delta T_{\rm eff} = 250$ &  $\Delta$log\,$g = 0.5$ & 
$\Delta \xi_{\rm turb} = - 1.0$ \\ 
   & [K] & [cgs] & [km\,s$^{-1}$]\\
\hline \\ 
H\,{\sc i}  & $-0.10$ & $-0.20$ & $< 0.05$   \\
C\,{\sc i}  & $+0.08$ & $-0.06$ & $+0.07$  \\
N\,{\sc i}  & $-0.13$ & $+0.18$ & $+0.03$  \\  
O\,{\sc i}  & $-0.07$ & $+0.09$ & $+0.10$  \\
Na\,{\sc i} & $+0.27$ & $-0.27$ & $+0.06$  \\
Mg\,{\sc i} & $+0.29$ & $-0.26$ & $+0.13$  \\
Al\,{\sc i} & $+0.25$ & $-0.26$ & $+0.00$ \\
Si\,{\sc i/ii} & $+0.25/-0.06$ & $-0.25/+0.05$ & $+0.00/+0.33$ \\
S\,{\sc i}  & $+0.17$ & $-0.15$ & $+0.00$ \\
Ca\,{\sc i} & $+0.37$ & $-0.30$ & $+0.01$  \\
Sc\,{\sc ii} & $+0.12$ & $+0.13$ & $+0.22$  \\
Ti\,{\sc ii} & $+0.12$ & $+0.03$ & $+0.22$  \\
Cr\,{\sc ii} & $+0.03$ & $+0.07$ & $+0.16$  \\
Fe\,{\sc i/ii} & $+0.34/+0.03$ & $-0.28/+0.07$ & $+0.01/+0.08$  \\
Ni\,{\sc i} & $+0.34$ & $-0.29$ & $+0.00$ \\
Zn\,{\sc i} & $+0.31$ & $-0.26$ & $+0.05$  \\
Sr\,{\sc ii} & $+0.31$ & $-0.06$ & $+0.23$ \\
Y\,{\sc ii}  & $+0.18$ & $+0.00$ & $+0.11$  \\
Zr\,{\sc ii} & $+0.13$ & $+0.03$ & $+0.05$ \\
Ba\,{\sc ii} & $+0.41$ & $-0.17$ & $+0.26$  \\
La\,{\sc ii} & $+0.28$ & $-0.04$ & $+0.01$  \\
Ce\,{\sc ii} & $+0.25$ & $-0.03$ & $+0.01$ \\
\hline  \\
\end{tabular}

\end{table}


Compared to other R\,CrB stars, V854\,Cen 
is only mildly hydrogen-deficient.
Only the hot R\,CrB DY\,Cen with log\,$\epsilon_{\rm H} = 10.8$ 
has a higher H abundance (Jeffery \& Heber 1993).
The H abundance of V854\,Cen is log\,$\epsilon_{\rm H}=9.9$ for
C/He=10\% and 8.9 for C/He=1\% which is the spectroscopically determined
C/He value for DY\,Cen. 
The next least H-deficient R\,CrB star is V\,CrA with
log\,$\epsilon_{\rm H} = 8.0$ for an adopted C/He=1\%. 
Synthetic spectra of the H$\beta$ profile in V854\,Cen is
shown in Fig. \ref{f:hbeta}. 
A solar hydrogen abundance is clearly excluded since it would require
unreasonably low $T_{\rm eff}$ and log\,$g$.
V854\,Cen accentuates the anti-correlation 
between the H and Fe abundances 
found for R\,CrB and EHe stars (Heber 1986; Lambert et al. 1998),
which Sakurai's object also follows (Asplund et al. 1997b).

Nitrogen when considered as a [N/Fe] ratio is slightly less overabundant
in V854\,Cen ([N/Fe]=1.4)
than in the R\,CrB majority stars 
(mean of 14 stars [N/Fe] = 1.6).
The [N/Fe] ratio in V854\,Cen is roughly 
consistent with a complete conversion of the original
CNO nuclei in a slightly metal-poor star to N through CNO-cycling.
Some additional N may also have 
been produced subsequently by proton-capture on $^{12}$C
synthesized from He-burning.
The [O/Fe] ratio for V854\,Cen, which is greater than 
seen in the R\,CrB majority stars, would seem to
require additional production of O through He-burning.  

\begin{figure}[t]
\centerline{
\psfig{figure=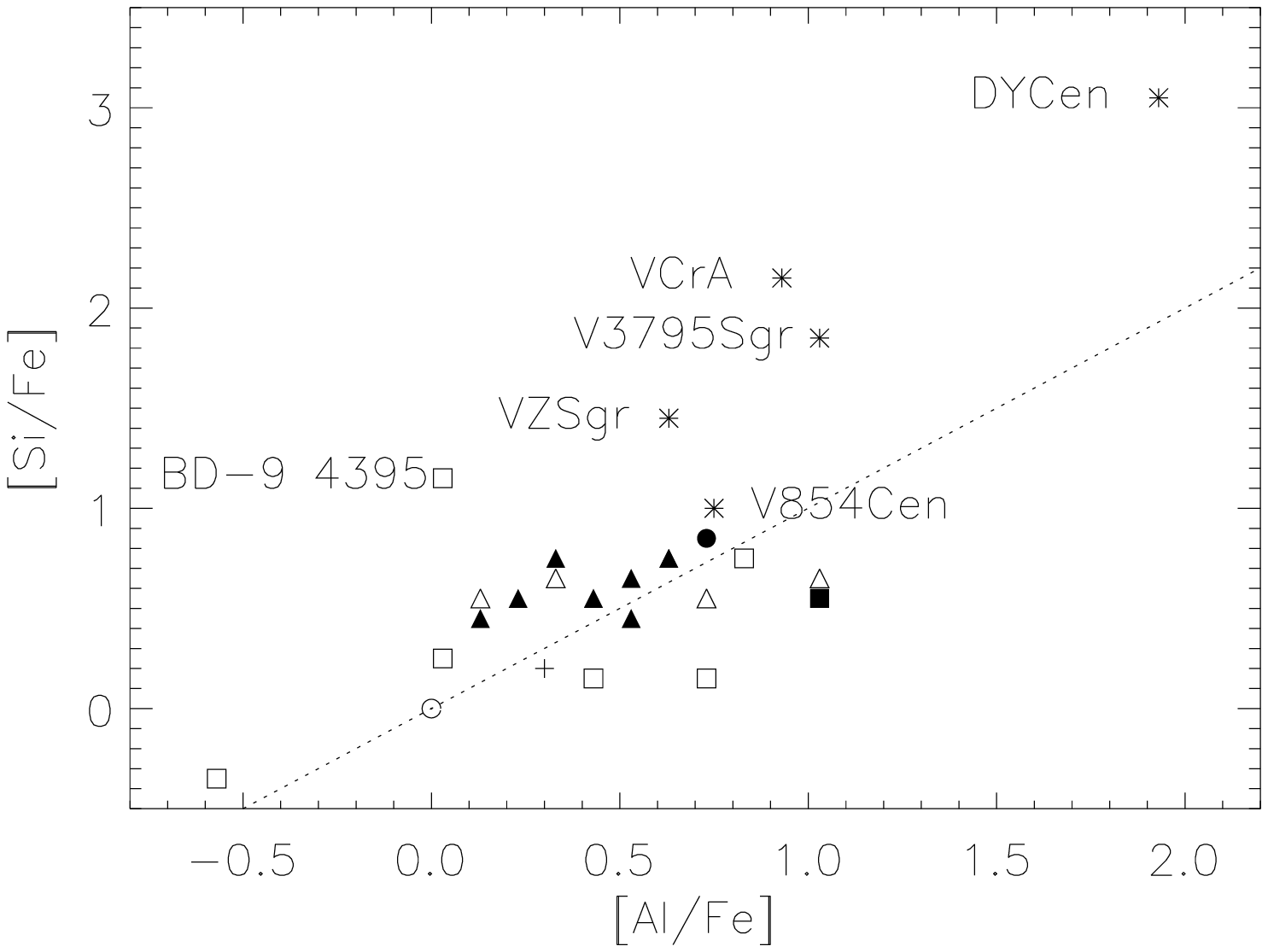,width=9.7cm}}
\caption{[Si/Fe] vs [Al/Fe] for H-deficient stars. 
The symbols correspond to Li-rich majority R\,CrB stars 
($\triangle$), other majority members (black triangles), 
the minority (including DY\,Cen) ($\ast$),
Lambert et al.'s (1998) EHe stars (black squares), 
Jeffery's (1996, see further references therein) 
`best' EHe stars ($\Box$),
Sakurai's object in October 1996 
($\bullet$) (Asplund et al. 1997b),  
the Sun ($\odot$) and typical
halo dwarf abundances for [Fe/H]\,$\simeq -1.0$ ($+$).
The dotted curve correspond to a 1-to-1 slope passing through
the solar values}
         \label{f:SiAl}
\end{figure}

[Na/Fe], [Al/Fe] (see Fig. \ref{f:SiAl}), 
[Si/Fe], [S/Fe], and to some degree [Ca/Fe],
are all overabundant relative to the Sun.
In particular, [Na/Fe]=1.6
is very high, which suggests that Na has
been synthesized through 
$^{22}$Ne(p,$\gamma$)$^{23}$Na;
of the analysed R\,CrB stars only V\,CrA has a higher [Na/Fe].
At the same time Al should have been
produced by $^{25}$Mg(p,$\gamma$)$^{26}$Al, 
but [Al/Fe] is not unusually high compared with 
the other R\,CrB stars. 
A possible explanation is that the proton captures
occurred in gas enriched in $^{22}$Ne. 
This is quite possible as CNO-cycling
converts all C, N, and O to $^{14}$N, which at higher temperatures is
converted by successive $\alpha$-captures to $^{22}$Ne. 
Ne is destroyed in
He-burning but, in a convective situation as may occur when H-rich
gas is mixed into the final He-shell, some may survive and be
available for conversion to Na. 
In steady conditions, Ne is converted to Mg in a He-shell.
Unfortunately the low $T_{\rm eff}$ prevents a determination
of the Ne abundance, but since [Mg/Fe] is only solar in V854\,Cen
the explanation seems plausible.

The ratios [Si/Fe]=1.0 and [S/Fe]=0.6 are higher than
expected for a dwarf star with the metallicity 
of V854\,Cen (i.e.[Si/Fe]\,$\simeq$\,[S/Fe]\,
$\simeq$\,0.2) but similar
to what has been determined for the R\,CrB majority.
Apparently, either Si and S have been synthesized or Fe has been
depleted (Lambert et al. 1998).
The slight Ca overabundance ([Ca/Fe]=0.2) is as expected 
for a mildly metal-poor dwarf (Edvardsson et al. 1993).

Of the Fe-group elements, Sc and Ti are overabundant while 
Cr has a solar abundance relative Fe: [Sc/Fe]=1.2, [Ti/Fe]=0.5
and [Cr/Fe]=0.0. In particular Sc is very overabundant, which suggests
synthesis by the $s$-process such that the Sc abundance is raised by
neutron captures on the much more abundant Ca nuclei.
This is supported by
the observed enhancements of the $s$-elements.
A similar phenomenon has been observed in the related stars
FG\,Sge (Acker et al. 1982; Gonzalez et al. 1998)
and Sakurai's object (Asplund et al. 1997b).
[Ti/Fe] is slightly higher than for metal-poor dwarfs, for which
[Ti/Fe]\,$\simeq 0.3$ is expected (Edvardsson et al. 1993),
though it could be due to observational errors.
Cr behaves similarly to Fe, as anticipated for a low metallicity star. 

\begin{table}[t]
\caption{Elemental abundance ratios in V854\,Cen compared to predictions
from $s$-processing calculations for different neutron exposures
$\tau_0$
\label{t:selement}
}
\begin{tabular}{lcccccc} 
\\
 \hline \\
Element & observed & \multicolumn {2} {c} {Single}  & & 
\multicolumn {2} {c} {Exponential}  \\
ratio   & ratio    & \multicolumn {2} {c} {exposure$^{\rm a}$} & & 
\multicolumn {2} {c} {exposure$^{\rm b}$}  \\
\cline{3-4} \cline{6-7} \\
        &          & $\tau_0 =0.1$  &  0.3 & & 0.05  & 0.1 \\
\hline \\
Ni/Fe   &  -0.1    &   -0.7       &   +0.3       & &  -1.0         & -0.7 \\
Zn/Fe   &  -1.6    &   -1.7       &   +0.2       & &  -1.9         & -1.3 \\
Sr/Fe   &  -3.8    &   -3.8       &   -1.2       & &  -3.6         & -2.2 \\
Y/Fe    &  -3.8    &   -4.7       &   -2.3       & &  -4.6         & -3.1 \\
Zr/Fe   &  -3.9    &   -4.6       &   -2.4       & &  -4.5         & -2.9 \\
Ba/Fe   &  -4.7    &   -4.6       &   -3.7       & &  -4.8         & -4.0 \\
La/Fe   &  -5.6    &   -5.8       &   -4.8       & &  -6.1         & -5.3 \\
Ce/Fe   &  -5.5    &   -5.6       &   -4.4       & &  -5.7         & -5.0 \\
\hline  \\
\end{tabular}

\begin{list}{}{}
\item[$^{\rm a}$] From Malaney (1987a).
\item[$^{\rm b}$] From Malaney (1987b).
\end{list}

\end{table}

Both [Ni/Fe]=1.1 and [Zn/Fe]=1.2 (see Fig. \ref{f:SiZn}) 
are distinctly non-solar, which 
cannot be attributed to an initial metal-poor composition for
V854\,Cen. Furthermore, the light $s$-process elements Y and Zr
are significantly enhanced ([Y/Fe]=1.4 and [Zr/Fe]=1.0)
and to lesser degree the heavy $s$-elements ([Ba/Fe]=0.6, [La/Fe]=0.7
and [Ce/Fe]=0.5),
which are all more abundant than for metal-poor dwarfs 
where [$s$/Fe]\,$\simeq 0.0$ is characteristic 
(Edvardsson et al. 1993). 
%
%
According to Malaney's (1987a) calculations of $s$-processing 
in a single exposure, the elements Ni-Ce suggest that V854\,Cen
has suffered a mild neutron exposure of $\tau_0 = 0.1 - 0.2$\,mb$^{-1}$, 
as shown in Table \ref{t:selement}. 
Since also Ni and Zn are well fit by the predictions, the stellar
atmosphere may consist predominantly of material exposed to
neutrons.  
If instead the elemental abundances are to be explained 
as the result of $s$-processing by
an exponential exposure, the derived abundances suggest 
$\tau_0 =0.05 - 0.1$\,mb$^{-1}$ (Malaney 1987b).
In this case, the observed [Ni/Fe]
is not well reproduced.
The estimated neutron exposure is similar to what seems appropriate
for most of the R\,CrB stars (Lambert et al. 1998), 
and indicates either that the formation of
an R\,CrB star in general 
produces an environment capable of mild $s$-processing,
or that the atmospheres have retained the $s$-process 
characteristics from the previous thermally pulsing AGB-phase. 
The latter possibility may likely be discounted as $s$--process enriched
AGB stars generally show a much more severe exposure to neutrons,
say $\tau_0 \approx 0.2 - 0.4$ mb$^{-1}$ (Busso et al. 1995). 
A final He-shell flash in a post-AGB star is clearly
able to produce these abundance patterns, 
as demonstrated by Sakurai's object.
It is more uncertain whether this is also possible in
the merger of two white dwarfs.

\subsection{Minority or majority status for V854\,Cen?}

The first survey of compositions of R\,CrB stars (Lambert \& Rao 1994) 
led to the definition of the two classes: majority and minority. The
latter were principally characterized by high [Si/Fe] and [S/Fe] ratios
and a low spectroscopic metallicity. 
The minority is also distinguished by their high [Na/Fe], 
[Al/Fe], [Ca/Fe] and [Ni/Fe] ratios.
\footnote{Rao \& Lambert (1996) placed V854\,Cen in the minority group
based on a preliminary  analysis that
overestimated the hydrogen abundance. 
As a result the continuous
opacity and resultant abundances were overestimated.  
In particular, the high [Si/Fe] suggested
a membership of the minority, but  the more accurate analysis
presented here  gives a lower [Si/Fe] ratio. 
The difference in [X/Fe] ratios 
is attributed to differences in temperature gradients
of the model atmospheres.
With the lower hydrogen content estimated here, 
the continuous opacity will be lower,
and hence backwarming will be more pronounced.} 

\begin{figure}[t]
\centerline{
\psfig{figure=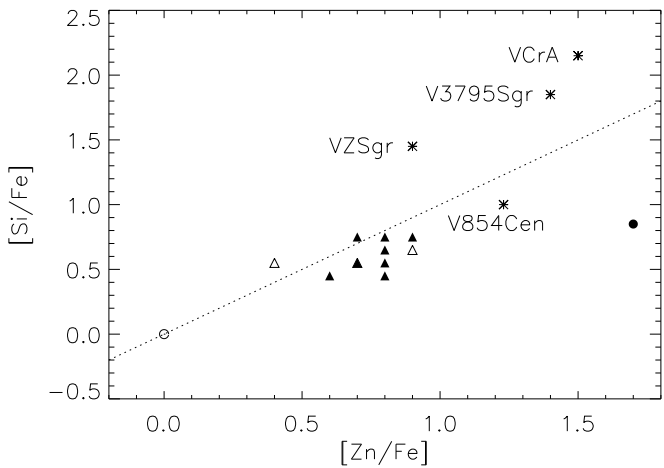,width=9.7cm}}
\caption{[Si/Fe] vs [Zn/Fe] for H-deficient stars. 
The symbols have the same meaning as in Fig. \ref{f:SiAl}}
         \label{f:SiZn}
\end{figure}

With the abundances determined here, V854\,Cen is mainly 
located in between
the three minority stars and the majority group, as shown in
e.g. Fig. \ref{f:SiAl}. 
Only in [Si/Fe] vs [Na/Fe] is V854\,Cen distinctly different from
the majority.
In particular, the [S/Fe] ratio is as expected for the majority
and far from the very high characteristic ratios of the minority. 
The high [Ni/Fe]=1.1 and [Zn/Fe]=1.2 (see
Fig. \ref{f:SiZn}) ratios 
are also atypical of the majority 
(on average 0.6 and 0.7, respectively) 
but typical of the minority for which both ratios show a large range.
The results for V854\,Cen may suggest that there is 
a gradual difference between 
the two groups introduced perhaps by
varying degree of dust depletion rather than reflecting
different evolutionary backgrounds. 

\subsection{Iron-depleted rather than iron-deficient?}

The peculiar abundances relative to Fe of V854\,Cen
may suggest that the star was not born as metal poor as its
Fe abundance indicates.
Dust depletion has been proposed to explain the observed 
abundance patterns in several post-AGB stars 
(cf. Bond 1991; Lambert 1996) and 
$\lambda$ Bo\"otis stars (Venn \& Lambert 1990), as well
as in the hot R\,CrB star DY\,Cen (Jeffery \& Heber 1993);
elements that condense readily into grains are now underabundant 
in these stellar photospheres.

\begin{figure}[t]
\centerline{
\psfig{figure=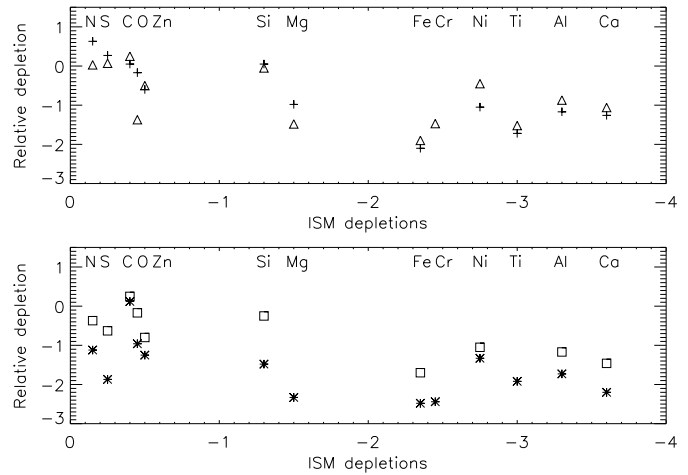,width=9cm}}
\caption{The depletions of the minority R\,CrB stars  vs the
observed ISM depletion (see text). 
The stellar depletion is here defined by
the stellar abundance relative to the solar abundance.
The symbols correspond to 
V3795\,Sgr ($\triangle$), V\,CrA ($+$), VZ\,Sgr ($\Box$) and 
V854\,Cen
($\ast$).
The different elements are indicated on top}
         \label{f:depletion}
\end{figure}

It is tempting to identify the low Fe abundance of the minority stars 
as the result of a dust-gas separation that either occurred in the post-AGB
progenitor or is occurring in the R\,CrB star. The temptation is
especially strong for V854\,Cen which is frequently in decline with
its surface obscured by dust.
A high [S/Fe] ratio is readily explained as S does not easily condense.
Moreover, a high [S/Fe] ratio is characteristic of those post-AGB stars
for which a severe separation of dust and gas has occurred.
However, the high [Si/Fe] is not na\"{\i}vely expected, 
in particular not if the dust-gas separation occurred 
in the C-rich gas of an R\,CrB star because in such an
environment a likely condensate is SiC. 
The [Si/Fe] ratio of the
extreme minority stars DY\,Cen, V\,CrA, and V3795\,Sgr greatly exceed 
the [Si/Fe] ratios seen in even those post-AGB stars 
most severely affected by the dust-gas separation.
In Fig. \ref{f:depletion} the observed depletions, defined here as
the stellar abundance relative to the solar value, are shown for
the minority members,
as a function of the observed depletions for the 
$\zeta$ Oph main cloud (Cardelli 1994). 
The latter cloud is taken as being representative of the ISM depletions.
Note that the observed depletions using this definition
do not necessarily reflect the exact amount of removed material,
since the initial metallicity may not have been solar,
but indicate the differential depletion between
different elements.
 
\begin{figure*}[t]
\centerline{
\psfig{figure=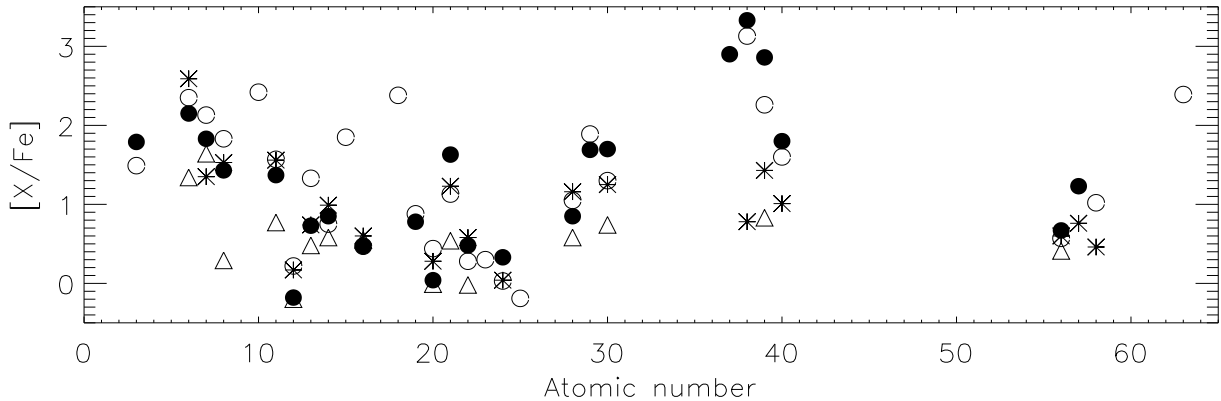,width=18cm}}
\caption{The elemental abundance ratios [X/Fe] 
in V854\,Cen ($\ast$) compared with Sakurai's object  
in May ($\circ$) and October ($\bullet$) 1996 (Asplund et al. 1997b)
and the mean of the majority R\,CrB stars 
($\triangle$) (Lambert et al. 1998).
Note that the apparent changes in [X/Fe] for Sakurai's object
between the two dates do not necessarily reflect the real alterations in
{\it absolute} abundances obvious in Table \ref{t:abund}, since
the abundance of Fe is not the same, which may or may not be real
since it is still within the observational uncertainties
}
         \label{f:abund}
\end{figure*}

Though there is some tendency for the abundances to follow
the ISM depletions as seen in 
Fig. \ref{f:depletion}, the correlation is not conclusive.
The differential depletion for the different elements 
for each star does not seem to exceed about 1.0\,dex.
Judging from Fig. \ref{f:depletion}, all elements, if 
depleted, seem to have been altered roughly by the same
amount, except possibly Si, S, Ni and Zn. 
In the case of S and Zn this might be expected but not
for Si and Ni.
A further complication is that the initial abundances of,
e.g., the $s$-process elements, such as Ni and Zn (see above), and 
the proton-capture elements like Na and Al, have likely
been modified by nucleosynthesis. 
This is exemplified by a slightly greater depletion of S
than Al in V854\,Cen, while in the ISM S reflects the initial metallicity
and Al being one of the most depleted elements.

Before drawing any definite conclusions more condensation 
calculations for H-deficient and C-rich environments
are needed. Such a special composition may well
cause significant changes in expected dust depletions 
compared to what is found in the H- and O-rich ISM.

\section{Abundance similarities with Sakurai's object}

In many respects V854\,Cen is similar to Sakurai's object,
which has probably recently experienced a  
final He-shell flash.
Sakurai's object is H-deficient and C-rich, like the R\,CrB stars.
Furthermore the abundances of some elements, most notably H, Li and the 
light $s$-process elements, seem to have
changed significantly within only five months in 1996,
presumably as a result of mixing of material exposed to 
nucleosynthesis with the stellar surface layers following
an ingestion of the H-rich envelope (Asplund et al. 1997b).
The abundances 
of Sakurai's object in May and October 1996 are listed
in Table \ref{t:abund}.
As shown in Fig. \ref{f:abund}, the differences 
in [X/Fe] between V854\,Cen and Sakurai's object in May and October 
1996 do not exceed 0.3\,dex for the elements studied
besides H, Li, N, and the $s$-process elements, and 
possibly C and Al.
As has already been mentioned, the H
abundances are also higher in these stars relative to
the R\,CrB stars in general.
Indeed, V854\,Cen is more similar to Sakurai's object than
any of the other R\,CrB stars, which suggests a 
similar evolutionary background as final flash objects
for the two stars.

Most of the abundance differences between the two stars 
could be interpreted 
as a result of less amount of H ingestion 
in connection with the He-shell flash.
The proton supply may not have been sufficiently high to raise the $^{13}$C 
abundance significantly (Renzini 1990), which in turn prevented 
$^{13}$C($\alpha$,n)$^{16}$O from being as efficient a neutron source
for $s$-processing as seems to have been the case in Sakurai's object.
Unfortunately, the C$_2$ Swan bands are too weak in V854\,Cen 
to admit a useful estimate of the $^{12}$C/$^{13}$C ratio 
but a high ratio would not be surprising.
More severe $s$-processing probably explains 
the higher [Sc/Fe] in Sakurai's object in October 1996 as well.
Li production is dependent on the amount of $^3$He in 
the envelope and on physical conditions, especially the 
convective velocity, at the time 
the Cameron-Fowler (1971) mechanism is initiated. 
It is not surprising that
two stars with otherwise similar compositions may differ in their
Li abundances.
The lower N abundance in V854\,Cen cannot, however,
be attributed to reduced CNO-cycling in connection with
the final flash, since the low observed $^{12}$C/$^{13}$C
and N/C ratios in Sakurai's object are not consistent with a 
significant simultaneous synthesis of N
(Asplund et al. 1997b).
The difference in [N/Fe] must therefore have been inherited
from earlier CNO-cycling and He-burning episodes.
As already noted, the derived [C/Fe] ratios are somewhat suspicious
in light of the ``carbon problem".

A possible problem with this interpretation is the slightly larger 
[Na/Fe] ratio in V854\,Cen by 0.2\,dex compared with Sakurai's object, 
which might indicate more proton captures. 
This minor difference
can presumably be attributed to uncertainties in the analyses
or that V854\,Cen might be of lower initial metallicity 
which would increase [Ne/Fe] and [Na/Fe].

It should be emphasized that Sakurai's object has continued to
evolve since October 1996, and its future composition may 
therefore not resemble V854\,Cen's. 
Our spectroscopic monitoring reveals 
that the star has become cooler in 1997, which is apparent from the
development of conspicuous molecular bands. The cooling
is also evident from photometry (Duerbeck et al. 1997).
The change in $T_{\rm eff}$ has probably caused an increased 
extension of the convective zone to deeper layers where 
further processed material could possibly be dredged up. 
This may, however, not necessarily imply 
a significant alteration of the elemental abundances, though
further studies of possible changes in the chemical composition
of the star is definitely encouraged.
We note that V854\,Cen resembles Sakurai's object more closely 
than any other R\,CrB star, both for May and October 1996, in 
spite of the modifications of the abundances during 1996. 
According to Fig \ref{f:abund}, some [X/Fe] ratios seem to evolve
towards the observed ratios in V854\,Cen, while other ratios
apparently become less similar. The latter are at least
partly due to a slight increase in the absolute abundance of Fe between the
two dates, which may or may not be real since it is 
still within the observational uncertainties.
Also, the significant amount of mixing of $s$-processed material 
that occurred
in Sakurai's object during 1996 explains the diverging ratios for the 
$s$-elements.
As already mentioned, the differences between V854\,Cen and 
Sakurai's object can largely be attributed to differing degree
of H-envelope ingestion. 
Therefore, the continued evolution of Sakurai's object, even
if it induces further abundance changes, does not
necessarily contradict
the possible interpretation of a similar background for the 
two stars.  
In particular, the resemblance between the two is based largely
on elements unlikely
to be modified by further nuclear processing and such 
similarities will likely still remain.

\section{Conclusions}

The derived chemical composition for V854\,Cen suggests that it
belongs to the minority group of the R\,CrB stars, though
less extreme than the other three minority members. One is tempted to
term it a minority-majority transition object. 
As an R\,CrB star, V854\,Cen is not only one of the stars with the
lowest Fe/C ratio but
also one of the least H-deficient. 
V854\,Cen accentuates the anti-correlation between the H and 
Fe abundances found for the R\,CrB and EHe stars
(Heber 1986; Lambert et al. 1998).
Like most R\,CrB stars, V854\,Cen
shows indications of a mild neutron exposure.

Perhaps, the most interesting result is the
abundance similarities between 
V854\,Cen and the final He-shell flash
candidate Sakurai's object (Asplund et al. 1997b). 
In fact, V854\,Cen resembles Sakurai's
object more closely than any of the other R\,CrB stars. This might 
suggest that the two stars also 
share the same evolutionary background.
Most of the abundance differences could if so be attributed to
less amount of H-envelope ingestion in connection with 
the final flash for V854\,Cen.

If V854\,Cen is indeed a final flash object, a search 
in optical or IR for a
fossil shell from a previous planetary nebula stage 
could be rewarding. 
In this context we note that nebular emission lines of [O\,{\sc i}],
[N\,{\sc ii}], [S\,{\sc ii}] and H$\alpha$, 
have already been observed
for the star during a decline (Rao \& Lambert 1993).
A connection between V854\,Cen and 
Sakurai's object would suggest that the final flash scenario 
is a viable channel to form R\,CrB stars.
Unfortunately, the status of V854\,Cen regarding 
class assignment as a minority or majority member is unclear.
It is therefore difficult to associate either group, if any, 
with being final flash objects.
An alternative interpretation is that all R\,CrB stars 
have been formed through the same mechanism 
but e.g. dust-gas separation
have introduced the differences in abundances between the two
groups, though such dust depletion must then have proceeded 
quite differently than in the ISM.
Further theoretical predictions for the two proposed models
regarding the resulting chemical compositions 
for different initial conditions are clearly
desirable in order to discriminate between them, as well
as work on the expected results of a dust depletion in H-deficient
and C-rich environments.
A continued monitoring of possible abundance changes in
Sakurai's object is clearly also a priority.

\begin{acknowledgements}
Nikolai Piskunov is thanked for help with the calculations of
hydrogen profiles.
The {\it gf}-values provided by Earle Luck are gratefully acknowledged.
This work has benefited from financial support from the 
Swedish Natural Research Council, Robert A. Welch Foundation 
of Houston, Texas, and National Science Foundation 
(grants AST9315124 and AST9618414). 
\end{acknowledgements}

\begin{table*}
\caption{Lines used for the abundance analysis. For the hydrogen and helium lines synthetic 
spectroscopy is used
\label{t:lines}
}
\begin{tabular}{lcrrr|lcrrr|lcrrr} 
 \hline 
 Species & $\lambda$  &  $\chi$ ~ & log\,gf  & $W_\lambda$~  &
 Species & $\lambda$  &  $\chi$ ~ & log\,gf  & $W_\lambda$~  &
 Species & $\lambda$  &  $\chi$ ~ & log\,gf  & $W_\lambda$~  \\ 
         &  [\AA]       & [eV]   &          &  [m\AA]       &
         &  [\AA]       & [eV]   &          &  [m\AA]       &
         &  [\AA]       & [eV]   &          &  [m\AA]       \\          
\hline 
H\,{\sc i}   & 4861.32 & 10.20 &  -0.02 &  synt & S\,{\sc i}   & 6757.16 &  7.87 &  -0.29 &  43 &Fe\,{\sc ii} & 4576.33 &  2.84 &  -3.04 & 191 \\
             & 4340.46 & 10.20 &  -0.45 &  synt & Ca\,{\sc i}  & 5590.12 &  2.52 &  -0.57 &  27 &             & 4582.84 &  2.84 &  -3.10 & 172 \\
He\,{\sc i}  & 5875.63 & 20.87 &   0.74 &  synt &              & 5598.49 &  2.52 &  -0.09 &  68 &             & 6084.10 &  3.20 &  -3.80 &  38 \\
Li\,{\sc i}  & 6707.80 &  0.00 &   0.17 & $<10$   &              & 5601.29 &  2.52 &  -0.52 &  48 &             & 6147.74 &  3.89 &  -2.74 &  86 \\
C\,{\sc i}   & 5813.51 &  8.87 &  -2.73 & 114   &               & 6122.22 &  1.89 &  -0.32 &  84 &              &6149.25 &  3.89 &  -2.75 &  92 \\
             & 5817.70 &  8.87 &  -2.86 &  75   &               & 6162.18 &  1.90 &  -0.09 & 100 &              &6247.55 &  3.89 &  -2.34 & 137 \\
             & 5850.25 &  8.77 &  -2.68 &  97   &               & 6166.44 &  2.52 &  -1.26 &  11 &              &6331.97 &  6.22 &  -1.64 &  17 \\
             & 5864.95 &  8.77 &  -3.55 &  57   &               & 6439.08 &  2.52 &   0.39 &  81 &              &6369.46 &  2.89 &  -4.36 &  32 \\
             & 5877.31 &  8.77 &  -2.14 & 104   &               & 6449.81 &  2.52 &  -0.50 &  23 &              &6432.68 &  2.89 &  -3.58 &  71 \\
             & 5963.99 &  8.64 &  -2.64 & 133   &               & 6462.57 &  2.52 &  -0.26 &  66 &              &6442.95 &  5.55 &  -2.46 &  16 \\
             & 5969.33 &  7.95 &  -3.08 & 110   &               & 6493.78 &  2.52 &  -0.39 &  30 &              &6446.39 &  6.22 &  -1.99 &   8 \\
             & 6292.37 &  9.00 &  -2.19 & 186   &  Sc\,{\sc ii} & 4354.61 &  0.60 &  -1.56 & 239 &              &6516.08 &  2.89 &  -3.29 & 136 \\
             & 6335.70 &  8.77 &  -2.80 &  87   &               & 4431.37 &  0.60 &  -1.88 & 170 &Ni\,{\sc i}   &5578.73 &  1.68 &  -2.74 &  25 \\
             & 6337.18 &  8.77 &  -2.33 & 137   &  Ti\,{\sc ii} & 4350.83 &  2.06 &  -1.81 & 134 &              &5592.28 &  1.95 &  -2.57 &  20 \\
             & 6342.32 &  8.77 &  -2.11 & 142   &               & 4450.49 &  1.08 &  -1.45 & 281 &              &5831.61 &  4.17 &  -0.96 &  10 \\
             & 6378.79 &  8.77 &  -3.29 &  76   &               & 4488.32 &  3.12 &  -0.82 & 196 &              &6643.64 &  1.68 &  -2.48 &  43 \\
             & 6578.77 &  9.00 &  -2.60 & 143   &               & 4493.53 &  1.08 &  -2.83 & 100 &              &6767.78 &  1.83 &  -2.17 &  49 \\
             & 6586.27 &  9.00 &  -2.02 & 152   &               & 4501.27 &  1.12 &  -0.75 & 340 &              &6772.32 &  3.66 &  -0.98 &  18 \\
             & 6591.45 &  8.85 &  -2.41 & 110   &               & 4518.30 &  1.08 &  -2.64 & 123 &Zn\,{\sc i}   &4810.53 &  4.08 &  -0.17 & 122 \\
             & 6595.24 &  8.85 &  -2.41 &  90   &               & 4529.46 &  1.57 &  -2.03 & 169 &              &6362.00 &  5.79 &   0.27 &  64 \\
             & 6611.35 &  8.85 &  -1.84 & 195   &               & 4544.01 &  1.24 &  -2.40 & 145 &Sr\,{\sc ii}  &4215.52 &  0.00 &  -0.16 & 460 \\
             & 6641.96 &  9.03 &  -3.46 &  39   &               & 4545.14 &  1.13 &  -2.46 & 157 &Y\,{\sc ii}   &4398.01 &  0.13 &  -1.00 & 240 \\
             & 6650.97 &  8.85 &  -3.52 &  45   &               & 4563.76 &  1.22 &  -0.96 & 330 &              &4682.32 &  0.41 &  -1.51 & 156 \\
             & 6671.82 &  8.85 &  -1.66 & 217   &               & 4568.31 &  1.22 &  -3.03 & 112 &              &4786.58 &  1.03 &  -1.29 & 134 \\
             & 6817.95 &  9.17 &  -2.43 & 105   &               & 4571.97 &  1.57 &  -0.53 & 382 &              &4823.31 &  0.99 &  -1.12 & 186 \\
             & 6828.12 &  8.54 &  -1.51 & 229   &               & 4589.96 &  1.24 &  -1.62 & 220 &              &5473.40 &  1.74 &  -1.02 & 123 \\
N\,{\sc i}   & 6440.94 & 11.76 &  -1.14 &   5   &               & 4779.99 &  2.05 &  -1.37 & 207 &              &5497.42 &  1.75 &  -0.58 & 176 \\
             & 6484.80 & 11.76 &  -0.76 &  15   &               & 4805.10 &  2.06 &  -1.10 & 249 &              &5521.56 &  1.74 &  -0.91 &  61 \\
             & 6644.96 & 11.76 &  -0.88 &  14   &               & 4874.02 &  3.10 &  -0.79 & 166 &              &5662.93 &  1.94 &   0.20 & 232 \\
             & 6793.84 & 11.84 &  -1.11 &   6   &               & 4911.20 &  3.12 &  -0.34 & 214 &              &5728.89 &  1.84 &  -1.12 &  54 \\
O\,{\sc i}   & 5958.48 & 10.99 &  -0.87 &  88   &  Cr\,{\sc ii} & 4275.57 &  3.86 &  -1.52 & 120 &              &6613.73 &  1.74 &  -1.10 &  87 \\
             & 6156.77 & 10.74 &  -0.44 & 135   &               & 4558.66 &  4.07 &  -0.45 & 219 &              &6795.41 &  1.73 &  -1.55 &  48 \\
             & 6158.18 & 10.74 &  -0.29 & 179   &               & 4588.22 &  4.07 &  -0.63 & 180 &Zr\,{\sc ii}  &4317.32 &  0.71 &  -1.38 &  50 \\
             & 6363.79 &  0.02 & -10.25 &  31   &               & 4616.05 &  4.07 &  -1.22 & 103 &              &4359.74 &  1.23 &  -0.56 &  94 \\
Na\,{\sc i}  & 5682.63 &  2.10 &  -0.71 & 152   &               & 4824.08 &  3.87 &  -1.22 & 185 &              &4379.78 &  1.53 &  -0.35 & 128 \\
             & 5688.20 &  2.10 &  -0.40 & 191   &               & 4876.37 &  3.86 &  -1.46 & 154 &              &4403.36 &  1.18 &  -1.12 &  57 \\
             & 6154.22 &  2.10 &  -1.57 &  50   &  Fe\,{\sc i}  & 5569.62 &  3.42 &  -0.49 &  70 &              &4440.45 &  1.21 &  -1.19 &  66 \\
             & 6160.75 &  2.10 &  -1.27 &  75   &               & 5586.77 &  3.37 &  -0.10 & 104 &              &4454.80 &  0.80 &  -1.35 & 132 \\
Mg\,{\sc i}  & 4702.99 &  4.33 &  -0.38 & 166   &               & 5569.63 &  3.42 &  -0.49 &  70 &              &4494.41 &  2.40 &   0.07 & 114 \\
Al\,{\sc i}  & 6696.03 &  3.14 &  -1.32 &  13   &               & 6065.49 &  2.61 &  -1.53 &  42 &              &4495.44 &  1.20 &  -1.42 &  42 \\
             & 6698.67 &  3.14 &  -1.62 &   5   &               & 6136.62 &  2.45 &  -1.40 &  47 &              &4496.97 &  0.71 &  -0.81 & 163 \\
Si\,{\sc i}  & 5708.41 &  4.95 &  -1.47 &  46   &               & 6137.70 &  2.59 &  -1.40 &  36 &              &6114.78 &  1.66 &  -1.48 &   7 \\
             & 5772.26 &  5.08 &  -1.78 &  24   &               & 6230.74 &  2.56 &  -1.28 &  40 &Ba\,{\sc ii}  &4554.04 &  0.00 &   0.17 & 288 \\
             & 6091.92 &  5.87 &  -1.40 &   8   &               & 6252.55 &  2.45 &  -1.69 &  18 &              &5853.68 &  0.60 &  -1.00 & 134 \\
             & 6125.03 &  5.61 &  -1.51 &  29   &               & 6430.85 &  2.18 &  -2.01 &  20 &              &6141.73 &  0.70 &  -0.08 & 276 \\
             & 6145.02 &  5.61 &  -1.48 &  17   &  Fe\,{\sc ii} & 4351.76 &  2.71 &  -2.10 & 250 &              &6496.90 &  0.60 &  -0.38 & 250 \\
Si\,{\sc ii} & 6347.09 &  8.12 &   0.26 & 324   &               & 4508.23 &  2.86 &  -2.21 & 263 &La\,{\sc ii}  &4322.51 &  0.17 &  -1.05 &  17 \\
             & 6371.36 &  8.12 &  -0.05 & 237   &               & 4522.63 &  2.84 &  -2.03 & 306 &              &4429.90 &  0.24 &  -0.37 &  42 \\
S\,{\sc i}   & 5706.11 &  7.87 &  -0.80 &  21   &               & 4541.52 &  2.86 &  -3.05 & 174 &Ce\,{\sc ii}  &4460.21 &  0.48 &   0.17 &  35 \\
             & 6743.58 &  7.86 &  -0.70 &  23   &               & 4555.80 &  2.83 &  -2.29 & 264 &              &4562.37 &  0.47 &   0.32 &  24 \\
             & 6748.79 &  7.86 &  -0.44 &  33 \\
\hline  \\
\end{tabular}

\end{table*}

 \end{document}